\documentclass[reprint,amsmath,amssymb,aps,nolongbibliography,superscriptaddress]{revtex4-2}

\usepackage{graphicx} 
\usepackage{latexsym}
\usepackage{amsmath}
\usepackage{wasysym}
\usepackage{nicefrac}
\usepackage{color}
\usepackage{dcolumn}
\usepackage{bm}
\usepackage{footmisc}
\usepackage{braket}
\usepackage[colorlinks=true,citecolor=blue,urlcolor=blue,linkcolor=blue]{hyperref}

\newcommand{\dagga}{{\phantom{\dagger}}}

\begin{document}

\title{Influence of the inter-orbital interaction and kinetic terms on superconductivity: a simple two-orbital Hubbard model}

\author{Vito Marino}
\affiliation{SISSA, International School for Advanced Studies, Via Bonomea 265, I-34136 Trieste, Italy}

\author{Diego Florez-Ablan}
\affiliation{SISSA, International School for Advanced Studies, Via Bonomea 265, I-34136 Trieste, Italy}

\author{Luca F. Tocchio}
\affiliation{Institute for Condensed Matter Physics and Complex Systems, DISAT, Politecnico di Torino, I-10129 Torino, Italy}

\author{Massimo Capone}
\affiliation{SISSA, International School for Advanced Studies, Via Bonomea 265, I-34136 Trieste, Italy}

\author{Federico Becca}
\affiliation{Dipartimento di Fisica, Università di Trieste, Strada Costiera 11, I-34151 Trieste, Italy}

\date{\today}

\begin{abstract}
We investigate a minimal two-orbital Hubbard model with intra- and inter-orbital nearest-neighbor hopping $t$ and $\tilde{t}$, as well as intra- and inter-orbital density-density interactions $U$ and $U'$ by means of variational {\it ans\"atze} based on Jastrow-Slater wave functions within quantum Monte Carlo techniques. To focus on the electronic mechanisms of superconductivity, we restrict the variational {\it ans\"atze} to uniform nonmagnetic states with an explicit pairing amplitude and compute the pairing correlations as a function of filling and model parameters. At $U/t=10$, superconducting correlations are highly enhanced by the presence of inter-orbital terms, $U'$ and  $\tilde{t}$. For $\tilde{t}=0$, a finite value of $U'$ effectively screens the intra-orbital repulsion $U$, producing a shift in the superconducting dome. Consequently, inter-orbital repulsion yields a sizable increase in electron pairing compared to the single-orbital baseline. Furthermore, introducing a finite $\tilde{t}$ provides an additional boost to superconducting correlations, an effect driven by the simultaneous presence of flat and broad bands in the electronic structure.
\end{abstract}

\maketitle

\section{Introduction}\label{sec:intro}

The emergence of superconductivity from purely repulsive electron-electron interactions remains one of the central problems in contemporary condensed-matter physics. In particular, unconventional superconductivity is frequently found in close proximity to a Mott insulating phase, suggesting that strong local correlations can actively enhance pairing tendencies rather than simply suppress charge motion. This paradigm shift was heavily driven by Anderson~\cite{anderson1987}, who proposed that high-temperature superconductivity may naturally emerge from a Mott insulator, described in terms of a resonating-valence-bond (RVB) state~\cite{fazekas1974}. In this framework, the preformed singlet pairs, inherent to a correlated Mott insulator, become mobile upon doping, effectively giving rise to a superconducting state directly out of a repulsive electronic background~\cite{dagotto1994,lee2006}. This mechanism fundamentally contrasts with conventional Bardeen-Cooper-Schrieffer (BCS) theory, transforming the Coulomb repulsion from a barrier to pairing into its primary driving force. 

While the single-orbital Hubbard model has provided a fundamental framework for understanding correlation effects and basic Mott physics, many materials possess multiple active orbitals near the Fermi level. In such cases, spin, charge, and orbital degrees of freedom are simultaneously active, giving rise to a rich competition between different instabilities and making the microscopic origin of superconductivity an even more challenging issue. The relationship between superconductivity and strong correlations in a multi-orbital (or multi-component) scenario can be relevant to a broad class of quantum materials, ranging from iron-based superconductors~\cite{kamihara2008,scalapino2012,georges2013,demedici2014,fernandes2022} to recently discovered Moir\'e materials~\cite{cao2018,balents2020} and nickelates~\cite{nomura2022}.

The simplest scenario, where inter-orbital terms can influence the enhancement or suppression of superconductivity, is a two-orbital Hubbard model on a square lattice. Starting from two decoupled single-orbital Hubbard models, each defined by intra-orbital nearest-neighbor hopping $t$ and interaction $U$, the coupling may be achieved through a potential term, represented by an on-site inter-orbital density-density interaction $U'$. The SU(4)-symmetric case with $U=U'$ has been extensively studied in the past with a focus on the metal-insulator transitions at integer filling~\cite{rozenberg1997,buenemann1998,koch1999,facio2017}. A similar model with opposite hopping for the two orbitals has been used to characterize excitonic phases in electron-hole bilayers~\cite{giuli2023}. 

We underline that our model does not include the Hund's coupling, which has been widely studied mainly by means of dynamical mean-field theory (DMFT)~\cite{georges2013,koga2015}. Our model is not meant to describe any specific material, but rather to provide a clean picture of fundamental effects introduced by the orbital degree of freedom in one of the simplest realizations. Indeed, in the attractive regime, this simplified density-density interaction has been shown to reproduce the essential superconducting-to-insulator phenomenology also obtained with the full density-density Kanamori Hamiltonian~\cite {torchia2025}.

Beyond the solid-state domain, ultracold fermionic alkaline-earth-like atoms provide a highly promising platform for the realization of a multi-orbital framework, like $^{173}$Yb~\cite{Yb_SUN} and $^{87}$Sr~\cite{Sr_SUN}. These systems offer exceptional control over Hamiltonian parameters and allow one to simulate clean systems without disorder and lattice displacements, thereby turning off the electron-phonon coupling~\cite{Bloch_ManyBody_review}. Loading these quantum gases in optical lattices with suitable geometry, one can simulate multiorbital physics realizing Hubbard models with full SU(N) symmetry~\cite{Gorshkov_SUN} or with a controlled symmetry breaking~\cite{tusi2022}. An even more direct way to simulate two-orbital models with density-density interactions is to populate two different long-lived electronic configurations. Here, besides the intra-orbital Hubbard term, the on-site inter-orbital density-density interaction may be tuned by a Feshbach resonance~\cite{pagano2015,hofer2015}. Therefore, this setup represents a pristine simulator for studying how purely density-driven inter-orbital interactions influence pairing tendencies.

In the second part of this work, we consider the case where the two orbitals are also  coupled by direct inter-orbital hopping processes. While an on-site hybridization term immediately opens a gap that yields a trivial band insulator at half-filling, a nearest-neighbor hopping introduces a much more profound effect by introducing distinct kinetic energy scales.  Specifically, it splits the two  non-interacting bands, which were originally equivalent, into two bands whose dispersion relations are proportional to the sum and the difference of the inter- and intra-orbital hoppings, respectively, thus resulting in a flatter and a wider band. In this regard, (nearly) flat bands have been widely discussed as favorable platforms for superconductivity, both because of their enhanced density of states and, in topological settings, because of the geometric contribution to the superfluid weight~\cite{peotta2015,aleksi2016,torma2021}. Beyond purely topological mechanisms, multiband systems may also benefit from the coexistence of weakly and strongly dispersive electronic states, where pairing tendencies are enhanced by the former while phase stiffness can be supported by the latter~\cite{berg2008,salasnich2019,paramasivam2024}.

Understanding whether and how superconducting order emerges in proximity to a Mott transition in multi-orbital systems is therefore a fundamental problem with potential applications in different directions. It is important to clarify whether orbital degeneracy provides an additional channel for enhancing pairing correlations, or whether it instead favors competing localization effects that suppress superconductivity. In this work, we investigate the superconducting properties of the two-orbital model on a square lattice by means of the variational Monte Carlo (VMC) method, complemented by dynamical mean-field theory (DMFT). Our aim is not to construct the full phase diagram, including magnetic or charge-ordered phases, but rather to isolate the role of inter-orbital interactions and orbital degrees of freedom in controlling the evolution of superconductivity as a function of electron density. By explicitly computing pairing correlation functions on top of optimized correlated wave functions, we aim to establish whether superconducting order survives in the presence of different inter-orbital influences. We show that inter-orbital kinetic and/or potential terms may strongly enhance the superconducting correlations of the single-orbital Hubbard model (obtained within the same approach). In particular, at $U/t=10$, the superconducting dome is strongly affected by the inclusion of $U'$, leading to enhanced pairing correlations in the vicinity of half filling. This effect has a simple explanation in terms of the fact that double occupation of the same orbital is released when $U'$ increases, thus reducing the value of $U$ in an effective single-orbital model. Then, the optimal value of the electron pairing is reached. The inclusion of the inter-orbital nearest-neighbor hopping has an even larger effect on the electron pairing. However, in this case, a simple interpretation in terms of a single-orbital model cannot be obtained. In fact, even though a large contribution of the pairing comes from the electrons occupying the flat band, a considerable effect is due to inter-band terms. 

The paper is organized as follows: In Sec.~\ref{sec:model}, we introduce the model and the variational wave functions that we use. In Sec.~\ref{sec:level3}, we report our results; first, by including only the inter-orbital interaction $U'$, and then in the presence of the inter-orbital nearest-neighbor hopping $\tilde{t}$. Finally, in Sec.~\ref{sec:conclusion}, we summarize our main results.

\section{Model and method}\label{sec:model}

We consider a two-orbital Hubbard model with intra- and inter-orbital repulsions on the square lattice. The Hamiltonian is given by:
\begin{equation}\label{eq:Ham}
\mathcal{H} = \mathcal{H}_{t} + \mathcal{H}_{U} + \mathcal{H}_{U'}, 
\end{equation}
where
\begin{align}
\label{eq:hoppingham}
& \mathcal{H}_{t} = \sum_{R,d=x,y} \sum_{\sigma} \sum_{\alpha,\beta} t_{\alpha,\beta} c^\dagger_{R,\alpha,\sigma} c^\dagga_{R+d,\beta,\sigma} + \textrm{h.c.}, \\ 
& \mathcal{H}_{U} = U \sum_{R} \sum_{\alpha} n_{R,\alpha,\uparrow} n_{R,\alpha,\downarrow}, \\  
& \mathcal{H}_{U'} = U' \sum_R \sum_{\sigma,\sigma'} \sum_{\alpha<\beta} n_{R,\alpha,\sigma} n_{R,\beta,\sigma'},
\end{align}
where  $c^\dagger_{R,\alpha,\sigma}$ ($c^\dagga_{R,\alpha,\sigma}$) is the creation (annihilation) operator for fermions on site $R$, orbital $\alpha=1,2$, and spin $\sigma=\uparrow,\downarrow$. Here, a simple hopping structure is considered, with intra-orbital amplitudes $t_{11}=t_{22}=-t$, which set the energy scale (e.g., $t=1$) and an optional symmetric inter-orbital term $t_{12}=t_{21}=-\tilde{t}$. In addition, $n_{R,\alpha,\sigma}=c^\dagger_{R,\alpha,\sigma} c^\dagga_{R,\alpha,\sigma}$ is the electron density at site $R$ and orbital $\alpha$ for spin $\sigma$. The parameters $U$ and $U'$ regulate the intra-orbital and inter-orbital Coulomb interactions, respectively. For $U \neq U'$ and $\tilde{t} \neq 0$, the Hamiltonian exhibits SU(2) spin symmetry and U(1) charge symmetry. Additionally, the model is invariant under a $\mathbb{Z}_2$ orbital-exchange symmetry. We mention that, at the highly symmetric point $U'=U$ and $\tilde{t}=0$, the distinction between spin and orbital degrees of freedom disappears, resulting in a unified SU(4) symmetry.

The numerical results are obtained using VMC on $L\times L$ clusters (with $N=L^2$ sites) and $N_{e}$ electrons (giving a total density $n=N_{e}/N$). This approach may provide an accurate description of correlated systems through a variational {\it ansatz}, whose physical properties are easily evaluated within a Monte Carlo framework~\cite{becca2017}. In our implementation, the variational wave function consists of an uncorrelated Bardeen-Cooper-Schrieffer (BCS) state with singlet pairing, modified by a Jastrow factor that accounts for electron-electron correlations via both intra- and inter-orbital density-density terms:
\begin{equation}\label{eq:psi}
\ket{\Psi} = \mathcal{J}_{c} \ket{\Phi_0}.
\end{equation}
Here, $\ket{\Phi_0}$ is defined as the ground state of an auxiliary Hamiltonian that is quadratic in the electron operators:
\begin{equation}\label{eq:Ham_aux}
\mathcal{H}_{\rm aux} = \mathcal{H}_{0} + \mathcal{H}_{\rm BCS},
\end{equation}
where $\mathcal{H}_{0}$ contains the same hopping structure as in the original Hamiltonian of Eq.~\eqref{eq:hoppingham} and an orbital-dependent chemical potential:
\begin{equation}\label{eq:aux_kin}
\mathcal{H}_{0} = \mathcal{H}_{t} - \sum_{R,\alpha,\sigma} \mu_{\alpha} \; c^\dagger_{R,\alpha,\sigma} c^\dagga_{R,\alpha,\sigma},
\end{equation}
and $\mathcal{H}_{\rm BCS}$ contains BCS pairing terms between nearest-neighbor sites with $d$-wave symmetry:
\begin{equation}\label{eq:aux_BCS}
\mathcal{H}_{\rm BCS} = \sum_{R,d=x,y} \sum_{\sigma} \sum_{\alpha,\beta} \Delta^{\alpha,\beta}_{d,\sigma} c^\dagger_{R,\alpha,\sigma} c^\dagger_{R+d,\beta,{\bar \sigma}} + \textrm{h.c.},
\end{equation}
where the electron pairing satisfies $\Delta^{\alpha,\beta}_{d,\sigma}=-\Delta^{\alpha,\beta}_{d,{\bar \sigma}}$ (i.e., singlet pairing), with  $\Delta^{\alpha,\beta}_{x,\sigma}=-\Delta^{\alpha,\beta}_{y,\sigma}$ (i.e., $d$-wave symmetry). In the following, we take $\Delta^{1,1}_{d,\sigma}=\Delta^{2,2}_{d,\sigma}$ and $\Delta^{1,2}_{d,\sigma}=\Delta^{2,1}_{d,\sigma}$, by denoting $\Delta^{1,1}_{x,\uparrow}=\Delta$ and $\Delta^{1,2}_{x,\uparrow}=\tilde{\Delta}$, which represent the two BCS parameters that characterize the variational wave function.

In addition, $\mathcal{J}_{c}$ denotes the density Jastrow factor:
\begin{equation}
\mathcal{J}_{c} = \exp{ \Big( -\frac{1}{2} \sum_{R,R'} \sum_{\alpha,\beta}  v^{\alpha,\beta}_{R,R'} \, n_{R,\alpha} n_{R',\beta} \Big)},
\end{equation}
where $n_{R,\alpha}=\sum_{\sigma} n_{R,\alpha,\sigma}$ is the electron density on site $R$ and orbital $\alpha$. The presence of a long-range Jastrow factor, which includes both intra- and inter-orbital terms, is essential to capture a Mott transition in a multi-orbital system~\cite{capello2005,defranco2018}. The parameters $v^{\alpha,\beta}_{R,R'}$ represent the Jastrow pseudo-potentials that, together with $\Delta$, $\tilde{\Delta}$ and $\mu_{\alpha}$, are optimized to minimize the variational energy~\cite{becca2017,sorella2005}:
\begin{equation}\label{eq:varene}
E = \langle \mathcal{H} \rangle =\frac{\braket{\Psi|\mathcal{H}|\Psi}}{\braket{\Psi|\Psi}}.
\end{equation}
The minimum of the energy corresponds to the optimized variational wave function according to the variational principle. The variational energy can be written as a function of the different energy contributions, where the interaction terms depend on the number of ``doublons'' in the lattice, that is a generalization of the concept of double occupancy to a multi-orbital scheme:
\begin{equation}\label{eq:enpart}
E = E_{K} + E_{U} + E_{U'},
\end{equation}
where
\begin{align}
& E_{K} = \langle \mathcal{H}_{t} \rangle, \\
& E_{U} = \langle \mathcal{H}_{U} \rangle = U \sum_{R} \sum_{\alpha} \langle n_{R,\alpha,\uparrow} n_{R,\alpha,\downarrow}\rangle, \\
& E_{U'} = \langle \mathcal{H}_{U'} \rangle =U'\sum_R \sum_{\sigma,\sigma'} \sum_{\alpha<\beta} \langle n_{R,\alpha,\sigma} n_{R,\beta,\sigma'}\rangle.
\end{align}
Specifically, we define:
\begin{align}
\label{eq:D}
& D = \sum_{R} \sum_{\alpha} \langle n_{R,\alpha,\uparrow} n_{R,\alpha,\downarrow}\rangle, \\
\label{eq:Dprime}
& D' = \sum_R \sum_{\sigma,\sigma'} \sum_{\alpha<\beta} \langle n_{R,\alpha,\sigma} n_{R,\beta,\sigma'}\rangle,
\end{align}
which denote the total number of intra- and inter-orbital doublons, respectively.

After optimizing the variational parameters, the existence of non-zero BCS parameters does not automatically imply a genuine superconducting state. In fact, the Jastrow factor can suppress underlying correlations, so that the BCS parameters may not reflect the true physical order. This issue is particularly relevant in multi-orbital systems, where no one-to-one correspondence exists between pair correlations and BCS variational parameters, because of orbital hybridization~\cite{marino2025}. To reliably establish the presence of superconducting pairing, we therefore compute the singlet–singlet correlation functions as:
\begin{equation}
\mathcal{P}_{\alpha\beta}(r) = \frac{1}{N} \sum_{R} \langle P^\dagga_{R,\alpha\beta} P^\dagger_{R+rx,\alpha\beta} \rangle,
\end{equation}
where 
\begin{equation}
P^\dagga_{R,\alpha\beta} = c^\dagga_{R+y,\alpha,\downarrow} c^\dagga_{R,\beta,\uparrow} + c^\dagga_{R,\beta,\downarrow} c^\dagga_{R+y,\alpha,\uparrow},
\end{equation}
destroys a singlet made up of two electrons at nearest-neighbor sites (along $y$) belonging to orbitals $\alpha$ and $\beta$, respectively. Then, superconductivity exists whenever $\mathcal{P}_{\alpha\beta}(r)$ does not decay to zero at large distances in the thermodynamic limit. Specifically, we can define the square superconducting order parameter:
\begin{equation}\label{eq:SCOP}
\phi^2_{\alpha\beta} = \lim_{r \to \infty} \mathcal{P}_{\alpha\beta}(r),
\end{equation}
which, in practice, is evaluated at $r=L/2$ (due to periodic-boundary conditions).

In order to support our results, we also performed DMFT calculations at zero temperature with the adaptive sampling configuration interaction (ASCI) as the impurity solver~\cite{mejuto-zaera2019,florez2025}. For these calculations, we used 7 bath sites per orbital impurity, and we kept $\sim 10^5$ Slater determinants in the representation of the ground state wave function of the impurity model. The DMFT calculations are performed in the symmetric paramagnetic solution in which superconductivity is not allowed. 

\begin{figure}
\includegraphics[width=\columnwidth]{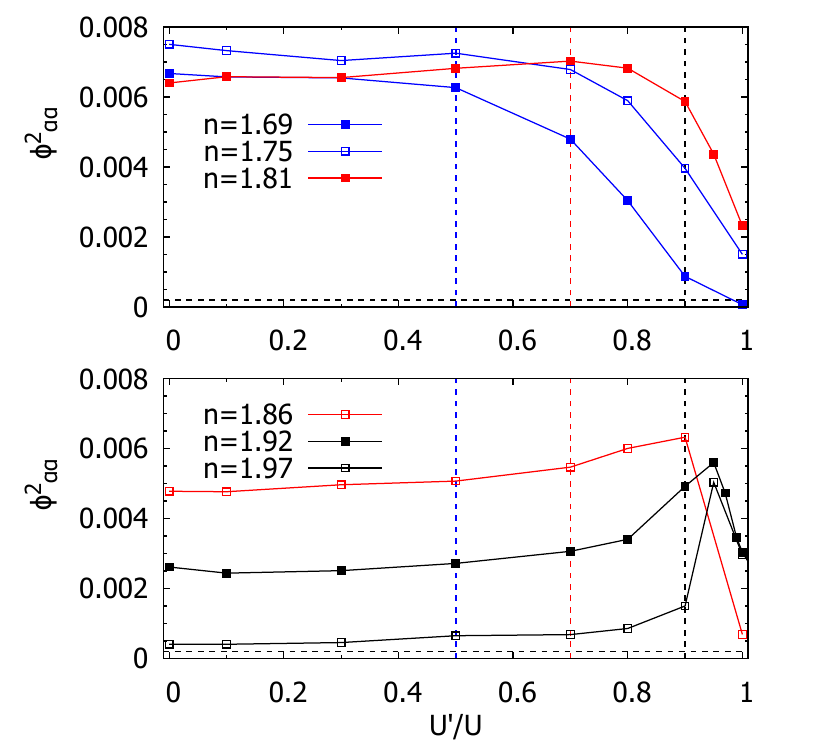}
\caption{\label{fig:OrdPar}
Superconducting order parameter of Eq.~\eqref{eq:SCOP} for $\alpha=\beta$ (e.g., for intra-orbital singlets) and different electron densities between $n=1.69$ and $n=2$. The horizontal dashed line indicates the  superconducting order parameter at $n=1.69$ in the noninteracting case ($U=U'=0$). The dashed vertical lines correspond to the values of $U'/U$ shown in  Fig.~\ref{fig:domeSC}. For clarity, the various electron densities are split in two panels: $n=1.69$, $1.75$, and $1.81$ (upper panel) and $n=1.86$, $1.92$, and $1.97$ (lower panel). Data are shown for the $L=12$ cluster and errorbars are smaller than the point size.}
\end{figure}

\begin{figure}
\includegraphics[width=\columnwidth]{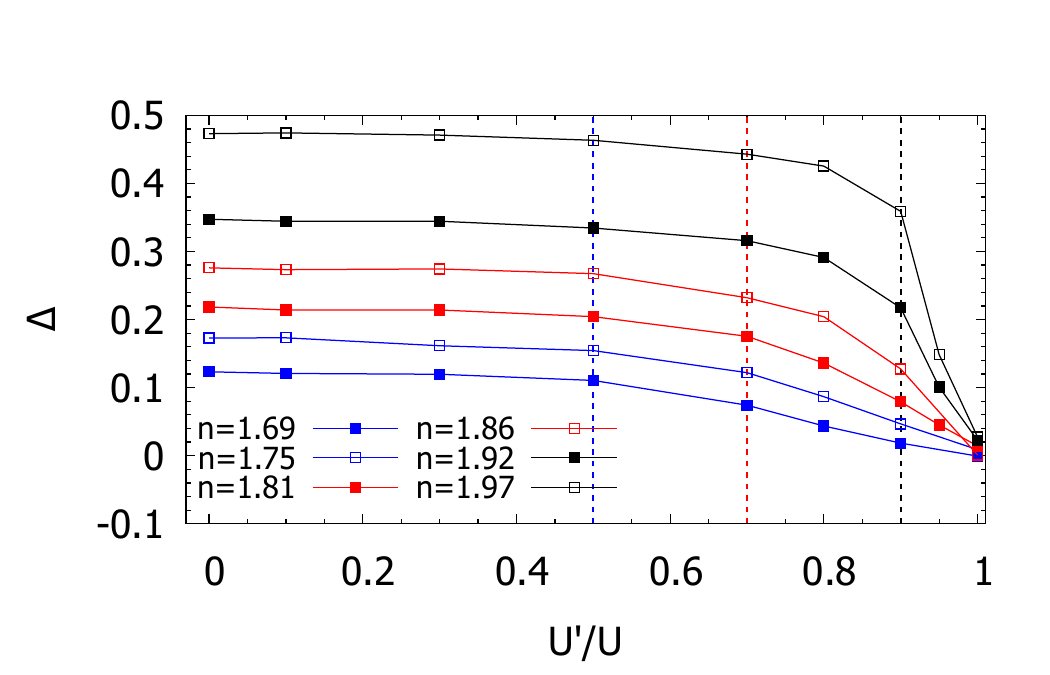}
\caption{\label{fig:varBCS}
Intra-orbital singlet BCS variational parameter $\Delta$ of Eq.~\eqref{eq:aux_BCS} as a function of $U'/U$ for different electron densities $n$ in a range from $1.69$ to $1.97$.}
\end{figure}

\section{Results}\label{sec:level3}

Here, we present our numerical results obtained via VMC. We begin by considering the decoupled case without inter-orbital hopping ($\tilde{t}=0$) to isolate the impact of the inter-orbital interaction $U'$ on the superconducting properties. Subsequently, we introduce a finite $\tilde{t}$ at fixed values of $U'$ to evaluate how inter-orbital hopping influences the electron pairing mechanism. In all cases, we focus our attention on the behavior of the pairing correlations when the electron doping is varied, by decreasing the density from $n=2$ (half filling) to $n=1.6$ (which corresponds to a relatively large hole concentration).

\subsection{The case with $\tilde{t}=0$}

In the following, the intra-orbital Coulomb interaction $U$ is fixed to a value larger than the bandwidth, namely $U/t=10$, and the inter-orbital term $U'$ is always smaller than $U$. In particular, for $U'=0$ the two orbitals are completely decoupled and the Hamiltonian is split into two equivalent single-orbital Hubbard models. In the latter case, the present approach, based on the variational wave function of Eq.~\eqref{eq:Ham_aux}, predicts a Mott insulator at half-filling~\cite{capello2006}; then, by reducing the electron density, $d$-wave superconductivity is established, with the typical dome-like behavior~\cite{eichenberger2007,yokoyama2013,misawa2014,tocchio2016}. In this case, the maximum superconducting signal is observed for $n \approx 1.75$. 

The presence of a finite $U'$ has an important effect, as observed in the intra-orbital pairing correlations. The results are shown in Fig.~\ref{fig:OrdPar}, where different electron densities are reported. The behavior is markedly different between small and large doping regimes (i.e., near and far from half filling, respectively). For example, for $n=1.69$, the superconducting order parameter smoothly decreases after an initial plateau, when moving from $U'=0$ to $U'/U=1$. A similar trend is also observed for $n=1.75$ and $1.81$, where the plateau progressively extends over a wider range of $U'$. Instead, close to half filling, e.g., for $n=1.86$, $1.92$, and $1.97$, after the initial plateau, the superconducting order parameter displays a rapid enhancement and develops a pronounced peak, followed by a dip for $U'/U=1$. Similarly to the single-orbital Hubbard model, the intra-orbital order parameter and the optimized variational BCS parameter $\Delta$ [see Eq.~\eqref{eq:aux_BCS}] do not exhibit the same trend (for $\tilde{t}=0$, the inter-orbital BCS parameter $\tilde{\Delta}$ is always vanishing or extremely small). In fact, the latter displays a smooth and monotonic behavior, decreasing as the electron density decreases and increasing as the inter-orbital Coulomb interaction increases, see Fig.~\ref{fig:varBCS}. In this context, we mention that the BCS parameter vanishes in the $U'/U=1$ point.

\begin{figure}
\includegraphics[width=\columnwidth]{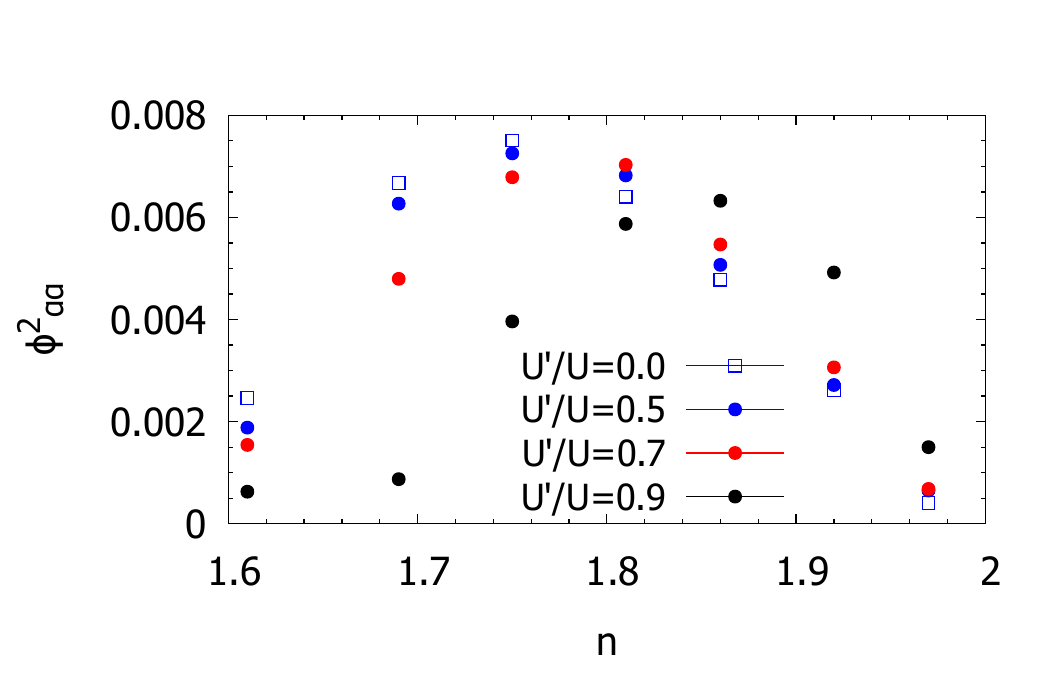}
\caption{\label{fig:domeSC}
Superconducting order parameter of Eq.~\eqref{eq:SCOP} for different values of $U'/U$ as a function of the electron density $n$. The full circles correspond to the vertical dashed lines in Fig.~\ref{fig:OrdPar}. The results for $U'=0$ are also reported for comparison. Data are shown for the $L=12$ cluster and errorbars are smaller than the point size.}
\end{figure}

\begin{figure}
\includegraphics[width=\columnwidth]{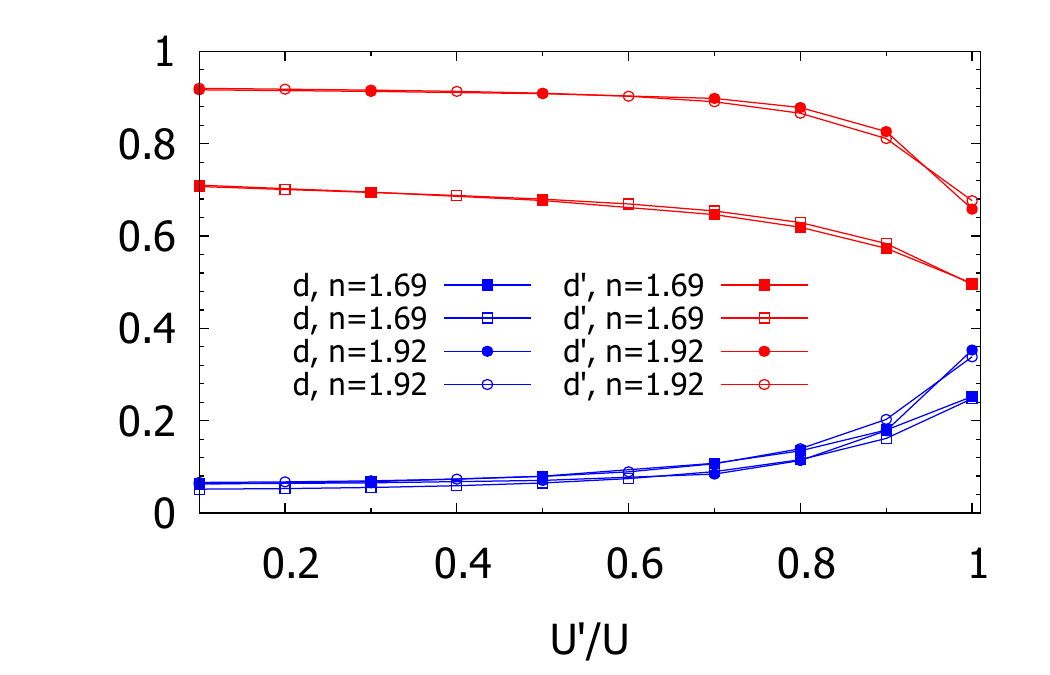}
\caption{\label{fig:ndUp}
Intra- and inter-doublon density $d=D/N$ and $d'=D'/N$ [see Eqs.~\eqref{eq:D} and~\eqref{eq:Dprime}] as a function of $U'/U$ for $n=1.69$ and $n=1.92$ computed by VMC (full symbols). The DMFT results (empty symbols) are also reported for comparison. The VMC calculations are performed on the $L=12$ cluster.}
\end{figure}

These results directly imply that the presence of the inter-orbital Coulomb interaction $U'$ shifts and squeezes the superconducting dome toward half filling, as shown in Fig.~\ref{fig:domeSC}. In fact, although the domes for $U'=0$ and $U'/U=0.5$ are nearly identical, there is a progressive shift towards half filling when larger values of the inter-orbital interaction are considered. Remarkably, despite a reduction in the maximum value of the dome as $U'/U$ increases from $0$ to $0.9$, the net effect is a significant enhancement of the pairing correlations near half filling.

A possible interpretation of this effect can be formulated in terms of an effective reduction of the intra-orbital repulsion $U$. Taking $U/t=10$ strongly suppresses configurations with intra-orbital double occupations, e.g., $D \approx 0$. For $U'=0$, the system decouples into two independent Hubbard models, where the inter-orbital double occupancy scales as $D' \approx N_{e}-N$. As $U'$ increases, configurations with finite $D'$ become also penalized, thus leading to an enhancement of the intra-orbital double occupations, which is equivalent to a reduction in the value of $U$. The crucial point is that the magnitude of the peak in the pairing correlations increases when the ratio $U/t$ is decreased down to $U/t \sim 6-8$, i.e, close to the Mott transition at half filling~\cite{tocchio2012}. This interpretation is supported by Fig.~\ref{fig:ndUp}, which shows the intra- and inter-orbital doublon densities ($d=D/N$ and $d'=D'/N$) as a function of $U'/U$, for $n=1.69$ and $1.92$. For $U'/U \lesssim 0.7$, the doublon populations remain essentially unchanged with respect to the $U'=0$ case, with $d' \gg d$. As $U'$ becomes comparable to $U$ (e.g., $U'/U \gtrsim 0.7$), a competing mechanism sets in, leading to an increase in $d$ and, as a consequence, a reduction in $d'$. Remarkably, our VMC results are in excellent agreement with DMFT calculations for both electron densities.

\begin{figure}
\includegraphics[width=\columnwidth]{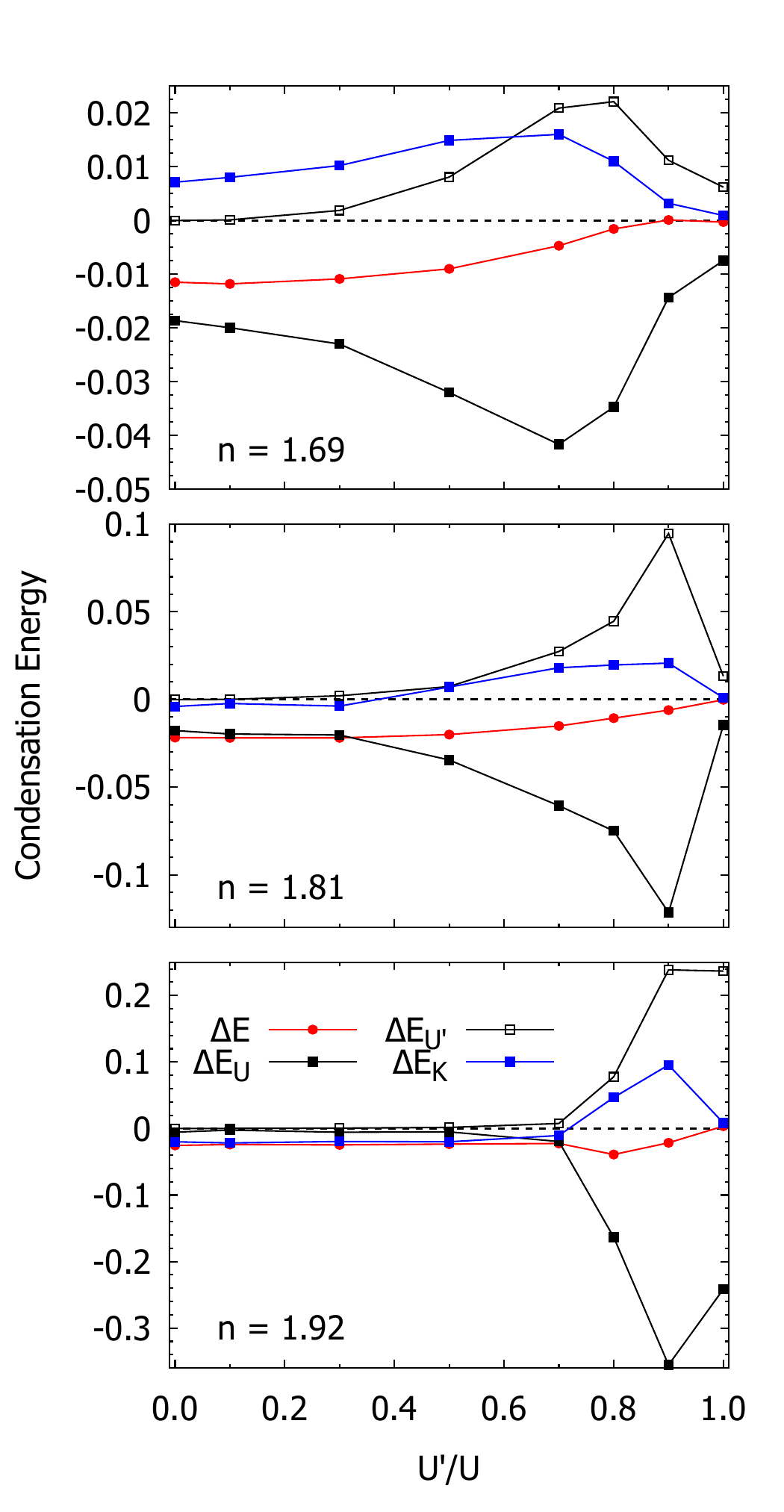}
\caption{\label{fig:conden}
Total superconducting condensation energy $\Delta E$ of Eq.~\eqref{eq:conden}. The various contributions, due to the intra-orbital Coulomb interaction ($\Delta E_U$), inter-orbital Coulomb interaction ($\Delta E_{U'}$), and kinetic energy ($\Delta E_{K}$), are shown.}
\end{figure}

Within this scenario, superconductivity is driven by a gain in the intra-orbital potential, while both kinetic and inter-orbital interactions oppose to the formation of electron pairing. In order to corroborate this fact, we consider the condensation energy:
\begin{equation}\label{eq:conden}
\Delta E = E_{\rm SC} - E_{\rm NonSC},
\end{equation}
where $E_{\rm SC}$ and $E_{\rm NonSC}$ are the variational energies [see Eq.~\eqref{eq:varene}] with and without including BCS pairing [see Eq.~\eqref{eq:aux_BCS}] in the auxiliary Hamiltonian of Eq.~(\ref{eq:Ham_aux}), respectively. The results for $\Delta E$, separated into three different contributions according to Eq.~\eqref{eq:enpart}, are shown in Fig.~\ref{fig:conden}. The energy gain always originates from the intra-orbital potential energy, since it decreases for any value of $U'/U$ when BCS pairing is included in the variational state. This energy gain is maximal for a value of $U'/U$ that depends on the electron density; in particular, at $n=1.92$ the energy gain due to the $\Delta E_U$ component of the condensation energy is maximal where the superconducting order is enhanced. By contrast, the contributions of kinetic energy and inter-orbital interaction are positive; the former remains nearly constant when $U'/U$ is varied, while the latter shows a clear increase when $U'$ approaches $U$. We also observe that the superconducting order is no longer energetically favorable for $U'/U=1$, where the condensation energy vanishes. This behavior supports the picture in which competing inter-orbital and intra-orbital interactions suppress pairing correlations at the SU(4)-symmetric point~\cite{defranco2018}. 

\subsection{The case with $\tilde{t} \ne 0$}

\begin{figure}
\includegraphics[width=\columnwidth]{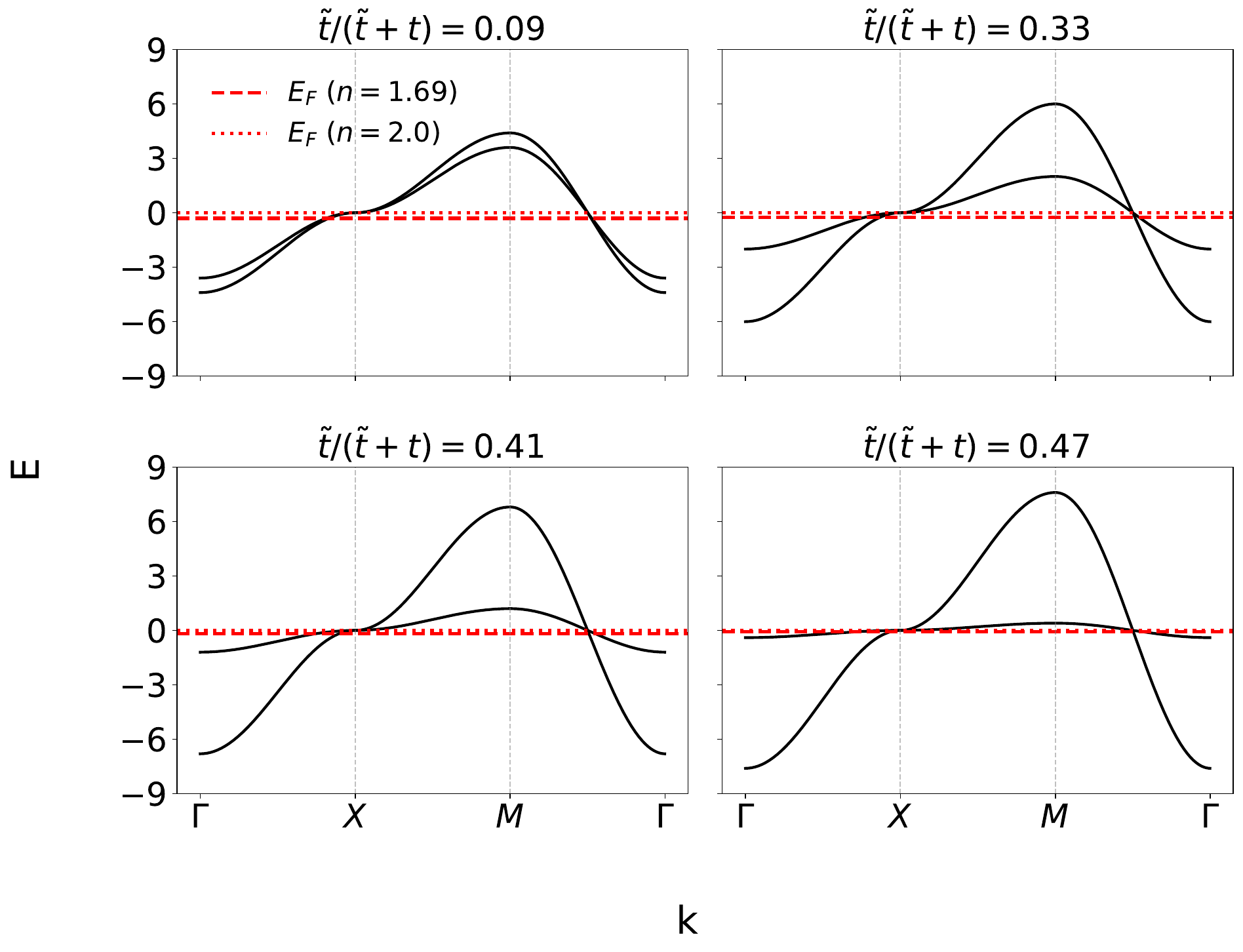}
\caption{\label{fig:bands}
Non-interacting bands obtained from Eq.~\eqref{eq:Hwithttilde}, for $\tilde{t}/t=0.1$ (upper-left panel), $0.3$ (upper-right panel), $0.7$ (lower-left panel), and $0.9$ (lower-right panel). The path in the BZ is $\Gamma \to X \to M \to \Gamma$, with $\Gamma=(0,0)$, $X=(\pi,0)$, and $M=(\pi,\pi)$. The Fermi energies for $n=1.69$ and $2$ are also shown.}
\end{figure}

The role of the inter-orbital degrees of freedom can be further enhanced by introducing an inter-orbital hopping term $\tilde{t}$, see the discussion below Eq.~\eqref{eq:Ham}. First of all, let us briefly discuss the non-interacting limit, described by:
\begin{equation}\label{eq:Hwithttilde}
{\cal H}_{t} = \sum_{k,\sigma} \epsilon_k
\begin{pmatrix}
c^\dagger_{k,1,\sigma}  & c^\dagger_{k,2,\sigma}
\end{pmatrix}
\begin{pmatrix}
t         & \tilde{t}  \\
\tilde{t} & t
\end{pmatrix}
\begin{pmatrix}
c^\dagga_{k,1,\sigma} \\
c^\dagga_{k,2,\sigma}
\end{pmatrix},
\end{equation}
where $\epsilon_k=-2(\cos k_k+\cos k_y)$, and $c^\dagger_{k,\alpha,\sigma}$ ($c_{k,\alpha,\sigma}$) are the creation (annihilation) operators in Fourier space. The diagonalization of the non-interacting Hamiltonian gives two bands with the same dispersion but different bandwidths, e.g., ${\cal E}_{k,\pm}=(t \pm \tilde{t}) \epsilon_k$. As $\tilde{t}\rightarrow t$, the bandwidth of the ${\cal E}_{k,+}$ band increases, whereas that of the ${\cal E}_{k,-}$ band flattens. A few representative cases with different values of $\tilde{t}$ are shown in Fig.~\ref{fig:bands}. The creation operators of the bands are simply written in terms of the ones for the orbitals, e.g., $c^\dagger_{k,\pm,\sigma} = \left ( c^\dagger_{k,1,\sigma} \pm c^\dagger_{k,2,\sigma} \right )/\sqrt{2}$. Then, the occupations of the broad and flat bands are then given by:
\begin{align}
\label{eq:nb}
& n_{b} =\frac{1}{N} \sum_{k,\sigma} \langle c^\dagger_{k,+,\sigma} c^\dagga_{k,+,\sigma} \rangle, \\
\label{eq:nf}
& n_{f} =\frac{1}{N} \sum_{k,\sigma} \langle c^\dagger_{k,-,\sigma} c^\dagga_{k,-,\sigma} \rangle.
\end{align}

\begin{figure}
\includegraphics[width=\columnwidth]{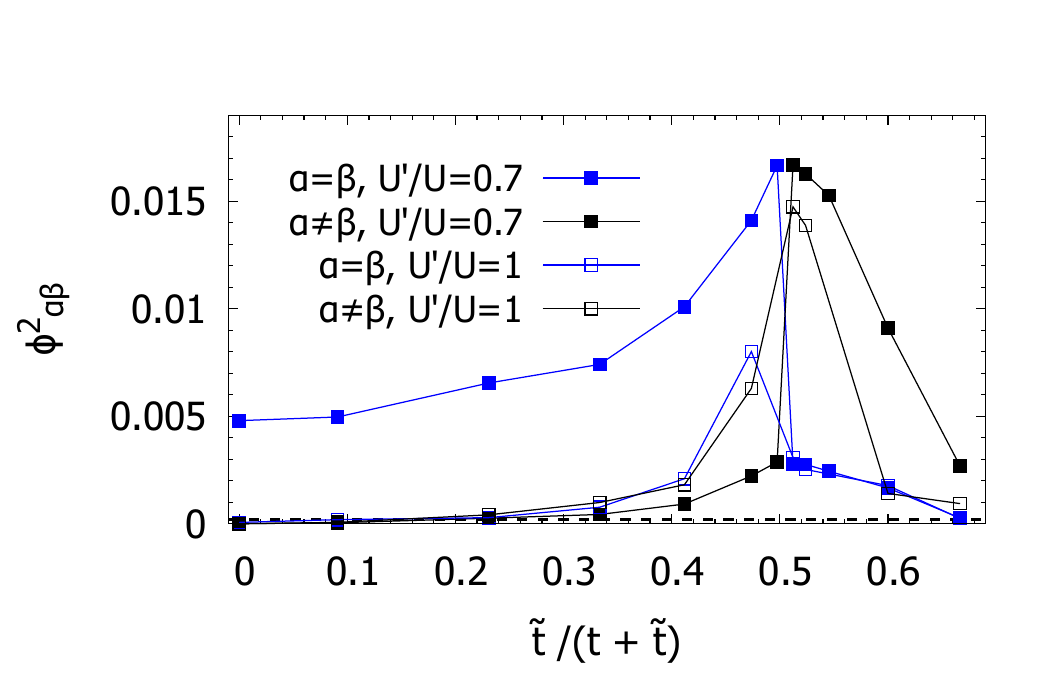}
\caption{\label{fig:interhop}
Superconducting order parameters resolved in the intra- ($\alpha=\beta$) and inter-orbital ($\alpha \neq \beta$) components in the presence of inter-orbital nearest-neighbor hopping $\tilde{t}$ at $n=1.69$ and $U/t=10$ for two different $U'/U$ ratios. Calculations are performed on the $L=12$ cluster. }
\end{figure}

In Fig.~\ref{fig:interhop}, we show the evolution of the $d$-wave superconducting order as a function of $\tilde{t}/(\tilde{t}+t)$, separating the intra-orbital contribution $\phi^2_{\alpha\alpha}$ and the inter-orbital one $\phi^2_{\alpha\beta}$ ($\alpha\neq\beta$), according to Eq.~\eqref{eq:SCOP}, at fixed $U/t=10$. Remarkably, a clear enhancement of the superconducting correlations is obtained by increasing $\tilde{t}/t$. In particular, the intra-orbital order parameter $\phi^2_{\alpha\alpha}$ increases up to $\tilde{t}/(\tilde{t}+t) \approx 0.5$, where it reaches its maximum value. For $\tilde{t}/(\tilde{t}+t) \gtrsim 0.5$ (i.e., $\tilde{t}\gtrsim t$) the inter-orbital component $\phi_{\alpha\beta}$ grows and eventually becomes the dominant superconducting channel, while $\phi_{\alpha\alpha}$ is suppressed. This behavior is observed for both $U'/U=0.7$ and $U'/U=1$, corresponding to superconducting and non-superconducting states when $\tilde{t}=0$, respectively. The increase in superconducting correlations is primarily associated to the enhancement of the intra-orbital BCS coupling $\Delta$ in the variational wave function.

\begin{figure}
\includegraphics[width=\columnwidth]{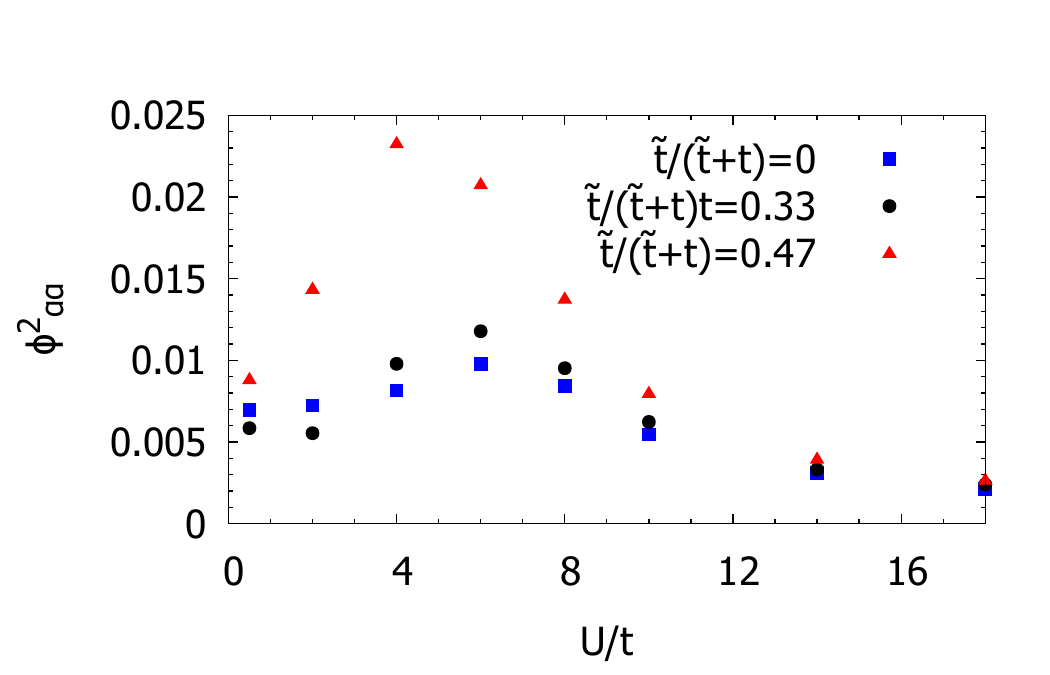}
\caption{\label{fig:interhop_U}
Intra-orbital superconducting order parameter $\phi^2_{\alpha\alpha}$ in the presence of inter-orbital nearest-neighbor hopping $\tilde{t}$ at $n=1.86$ and $U'/U=0.7$, as a function of $U/t$.}
\end{figure}

Interestingly, the maximum of $\phi_{\alpha\alpha}^2$ moves to lower values of $U/t$ as $\tilde{t}$ increases, see Fig.~\ref{fig:interhop_U}. This behavior is consistent with a reduction of the effective kinetic-energy scale associated with the progressive flattening of one of the bands, resulting in an enhanced pairing tendency at weaker interactions.

In the presence of $\tilde{t}$, a simple interpretation in terms of an effective single-band model, as suggested in the previous subsection for the effect of $U'$, is no longer possible. Indeed, in the regime where superconductivity is enhanced, the ground state remains strongly entangled both in the orbital and in the band representation, in which inter-band pairing correlations are finite (not shown). This prevents a simple description in terms of decoupled effective bands and the growth of $\phi^2_{\alpha\alpha}$ can be attributed to the coexistence of a quasi-flat band and a dispersive band in the pairing process. Since a finite $\tilde{t}$ hybridizes the two orbitals and promotes charge delocalization between them, Cooper pairs can gain kinetic energy from the broad band, owing to its larger dispersion, while the effective pairing strength remains enhanced because of the flat band, which reduces the effective ratio between bandwidth and Coulomb interaction. Finally, when the inter-orbital hopping exceeds the intra-orbital one, inter-orbital pairing processes become dominant, leading to a regime characterized by $|\tilde{\Delta}|>|\Delta|$. In the limit $\tilde{t}\gg t$, however, superconductivity is eventually suppressed, and both order parameters vanish.
 
\begin{figure}
\includegraphics[width=\columnwidth]{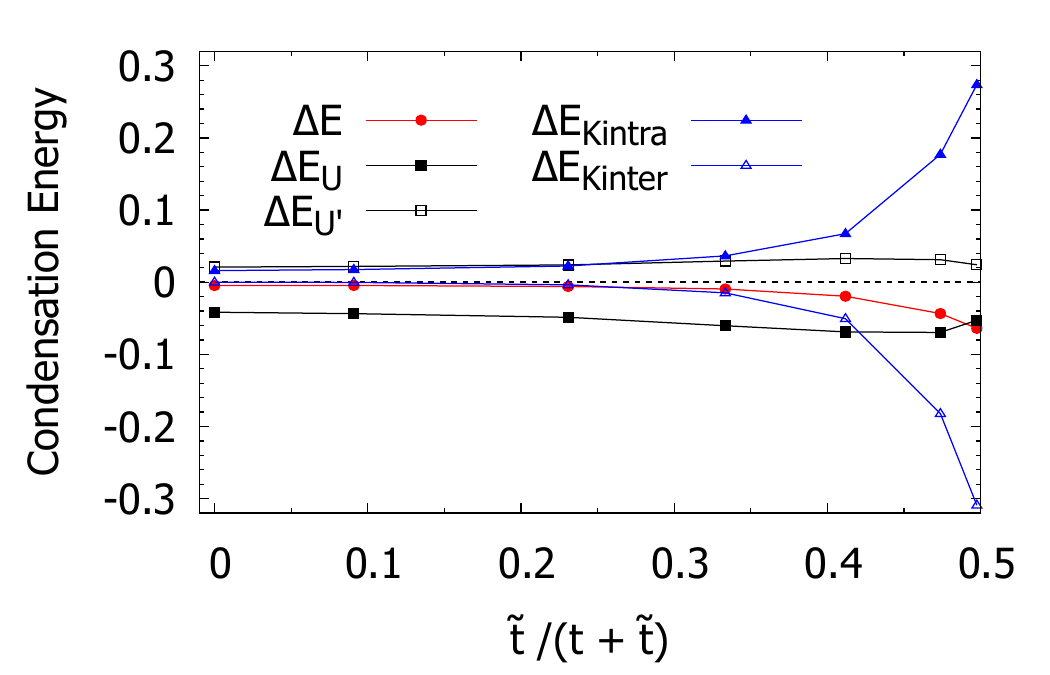}
\caption{\label{fig:condE_ttilde}
Total superconducting condensation energy $\Delta E$ of Eq.~\eqref{eq:conden} as a function of $\tilde{t}$. The various contributions, due to the intra-orbital Coulomb interaction ($\Delta E_U$), inter-orbital Coulomb interaction ($\Delta E_{U'}$), intra-orbital kinetic energy ($\Delta E_{Kintra}$), and inter-orbital kinetic energy ($\Delta E_{Kinter}$), are shown. Data are displayed for $n=1.69$, $U'/U=0.7$, and $U/t=10$ on the $L=12$ cluster.}
\end{figure}

Furthermore, we assess the energetic stability of the superconducting state in the presence of a finite inter-orbital hopping, as shown in Fig.~\ref{fig:condE_ttilde}. Here, we further split intra- and inter-orbital kinetic terms, to highlight their separate roles. The negative contribution to the condensation energy originates from the intra-orbital interaction term $\Delta E_U$, which remains almost constant as a function of $\tilde{t}$; in addition, the inter-orbital kinetic energy $\Delta E_{K_{\rm inter}}$ becomes predominant when increasing $\tilde{t}$, while the intra-orbital term displays the opposite effect. As a result, the total kinetic contribution, $\Delta E_{K_{\rm intra}}+\Delta E_{K_{\rm inter}}$, becomes negative before reaching $\tilde{t}/(t+\tilde{t}) = 0.5$, indicating that the kinetic-energy gain sets in as the system approaches the regime of strong inter-orbital hybridization.

\begin{figure}
\includegraphics[width=\columnwidth]{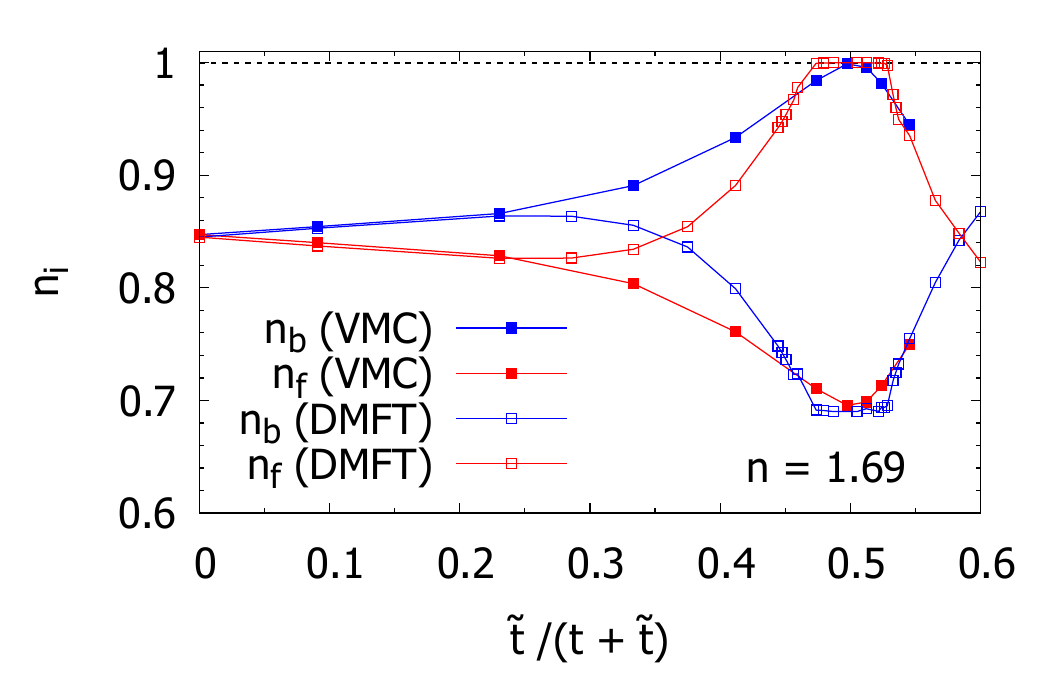}
\caption{\label{fig:nband_ttilde}
Comparison of the band occupation in VMC (full squares) and DMFT (empty squares) as a function of $\tilde{t}/(t+\tilde{t})$ at $n=1.69$, $U'/U=0.7$, and $U/t=10$. Here, $n_{b}$ and $n_{f}$ are the occupations of the broad and flat bands, respectively.}
\end{figure}

The presence of non-negligible BCS parameters is crucial to induce a redistribution of the electronic occupations $n_{b}$ and $n_{f}$ of the broad and flat bands [see Eqs.~\eqref{eq:nb} and~\eqref{eq:nf}], respectively. Within our VMC calculations  $n_{b}>n_{f}$, thereby favoring the kinetic-energy gain in the superconducting state, see Fig.~\ref{fig:nband_ttilde}. The enhancement of the superconducting order occurs while the flat band remains far from half filling, underscoring the crucial role played by this band in promoting superconductivity. Finally, we note that DMFT accurately reproduces the double occupations of VMC, which depend only weakly on $\tilde{t}$. In contrast, the agreement for individual band occupations is less satisfactory. For example, at $n=1.69$, VMC yields a monotonic trend in $n_{b}$ and $n_{f}$ up to $\tilde{t}/(t+\tilde{t}) = 0.5$, where $n_{b} \approx 1$. DMFT, on the other hand, reproduces the VMC results only at weak inter-orbital hoppings [i.e., $\tilde{t}/(t+\tilde{t}) \lesssim 0.25$]; beyond this point, the two occupations invert, and the flat band reaches half filling at $\tilde{t}/(t+\tilde{t}) = 0.5$. This behavior cannot be attributed solely to superconducting pairing, but also reflects the non-local correlations of the variational wave function.

\section{Conclusions}\label{sec:conclusion}

In this work, we have investigated the emergence the superconducting properties of a minimal two-orbital Hubbard model on the square lattice, focusing on the role of the inter-orbital Coulomb interaction $U'$ and of the inter-orbital hopping $\tilde{t}$. By using a Variational Monte Carlo approach with support from DMFT, we have evaluated the superconducting order parameter directly from long-distance pairing correlations, thus going beyond the information contained in the optimized BCS variational parameters alone. In the absence of inter-orbital hopping ($\tilde{t}=0$), we find that the inter-orbital interaction has a substantial effect on the superconducting correlations when $U'$ becomes comparable to the intra-orbital repulsion $U$. For small and intermediate values of $U'/U$, the superconducting order parameter is only weakly affected, especially at large hole doping. By contrast, close to half filling, the increase in $U'$ produces a significant enhancement of superconducting correlations before the symmetric point $U'/U=1$ is reached. This behavior can be interpreted as a displacement of the superconducting dome toward half filling, induced by the competition between intra-orbital and inter-orbital doublons. The analysis of the doublon densities provides a microscopic interpretation of this effect. When $U'$ is small, inter-orbital doublons are energetically inexpensive and dominate over intra-orbital doublons. As $U'$ approaches $U$, the value of the intra-orbital Coulomb repulsion is effectively varied by the presence of $U'$,  since inter-orbital double occupations are progressively penalized, while intra-orbital double occupations increase. This redistribution of local charge configurations effectively modifies the correlation strength felt by the system and shifts the optimal pairing regime. The agreement between the VMC and DMFT results for the doublon densities further supports this interpretation. The decomposition of the condensation energy clarifies the energetic origin of the superconducting state: The gain in energy mainly comes from the intra-orbital Coulomb contribution, while both the kinetic and inter-orbital interaction terms give positive contributions. In particular, the inter-orbital contribution becomes increasingly costly as $U'$ approaches $U$, eventually suppressing the energetic stability of superconductivity at the highly symmetric point. 

Finally, we have analyzed the effect of a finite inter-orbital hopping $\tilde{t}$. In the non-interacting limit, this term separates the spectrum into a more dispersive band and a progressively flatter band as $\tilde{t}\rightarrow t$. The coexistence of these two kinetic-energy scales enhances the superconducting correlations: electrons can gain kinetic energy through the dispersive band while the flat one changes the effective kinetic scale, strengthening the superconducting pairing. As a consequence, the intra-orbital superconducting order parameter is enhanced for values of $\tilde{t} \rightarrow t$, and the maximum of the superconducting dome is shifted to lower values of $U/t$. At $\tilde{t} > t$, inter-orbital pairing channels become increasingly relevant and dominate over the intra-orbital component, before superconductivity is eventually suppressed when the orbital hybridization becomes too strong. Overall, in the presence of a finite orbital hybridization, any mapping onto an effective single-band description becomes intrinsically ill-defined, since the orbital character is no longer conserved and the pairing correlations cannot be unambiguously assigned to independent bands or orbitals.

These findings highlight the importance of orbital fluctuations and inter-orbital processes in promoting or suppressing unconventional superconductivity in correlated multi-orbital systems and they can be used as a building block to understand superconducting tendencies in more realistic descriptions of multi-band systems. 

\begin{acknowledgments}
We would like to thank M. Fabrizio for useful discussions. We acknowledge financial support from the European Union–NextGenerationEU via the National Recovery and Resilience Plan PNRR MUR Project No. CN00000013-ICSC and MUR Project No. PE0000023-NQSTI. The views and opinions expressed are solely those of the authors and do not necessarily reflect those of the European Union, nor can the European Union be held responsible for them. High performance calculations were carried out thanks to resources provided by CINECA HPC via Convenzione SISSA 2025.
\end{acknowledgments}

\nocite{*}

\bibliography{apssamp}

\end{document}